\documentclass[epj]{svjour}

\usepackage{graphicx}
\usepackage{amstext}
\usepackage{amsbsy}
\usepackage{color}
\usepackage[tight,footnotesize]{subfigure}

\begin{document}

\title{Stabilizing spiral structures and population diversity in the asymmetric May--Leonard model through immigration}
\author{Shannon R. Serrao \and Uwe C. T\"auber}                    
\institute{Department of Physics \& Center for Soft Matter and Biological Physics, Virginia Tech \\
Robeson Hall, 850 West Campus Drive (MC 0435), Blacksburg, Virginia 24061, USA}

\date{Received: May 20, 2021 / Revised version: \today}

\abstract{
We study the induction and stabilization of spiral structures for the cyclic three-species stochastic May--Leonard model with asymmetric predation rates on a spatially inhomogeneous two-dimensional toroidal lattice using Monte Carlo simulations. 
In an isolated setting, strongly asymmetric predation rates lead to rapid extinction from coexistence of all three species to a single surviving population. Even for weakly asymmetric predation rates, only a fraction of ecologies in a statistical ensemble manages to maintain full three-species coexistence. 
However, when the asymmetric competing system is coupled via diffusive proliferation to a fully symmetric May--Leonard patch, the stable spiral patterns from this region induce transient plane-wave fronts and ultimately quasi-stationary spiral patterns in the vulnerable asymmetric region. 
Thus the endangered ecological subsystem may effectively become stabilized through immigration from even a much smaller stable region.
To describe the stabilization of spiral population structures in the asymmetric region, we compare the increase in the robustness of these topological defects at extreme values of the asymmetric predation rates in the spatially coupled system with the corresponding asymmetric May--Leonard model in isolation. 
We delineate the quasi-stationary nature of coexistence induced in the asymmetric subsystem by its diffusive coupling to a symmetric May--Leonard patch, and propose a (semi-)quantitative criterion for the spiral oscillations to be sustained in the asymmetric region.}


\authorrunning{S.R. Serrao, U.C. T\"auber}
\titlerunning{Stabilizing spiral structures and diversity in the asymmetric May--Leonard model}

\maketitle 

\section{Introduction}
The study of conditions for biodiversity and population extinctions is of course central to ecology and biology. 
For example, it is desirable that a malignant pathogen strain be rendered extinct in favor of a benign or dormant form; on the other hand, the survival and preservation of an endangered species within a larger ecosystem may be critical to the durability of the ecosystem as a whole. 
Consequently, theoretical and phenomenological models in ecology have sought to characterize and explore the coexistence of various populations in view of the relationships between the individuals that make up the whole environment~\cite{May73,Maynard74,Sigmund98,Murray02,Neal}.
Although the interactions are typically modelled between microorganisms, plants, and animals~\cite{Kingsland}, the theoretical tools developed here can be applied to (bio-) chemical reactions~\cite{Schuster2008UseOG}, genes~\cite{Hanski}, lasers~\cite{Kim2005ScalingBO}, economics~\cite{Malthus}, epidemics, and cancer growth~\cite{Vineis2006ThePD} as well, among other fields of current research.

Modelling schematic, idealized representations of the real world crucially relies on the underlying space structure in which they are realized. 
Models simulated on spatially extended systems produce phenomena that often cannot be adequately captured by a mere non-spatial or well-mixed (mean-field) variant. 
Indeed, spatial models for chemical reactions, competing populations, or spreading diseases feature a wide variety of activity fronts, oscillating structures, and propagating wave phenomena that differ widely in their dynamical properties~\cite{Cross,Cross09}.
In addition, stochastic fluctuations and spatio-temporal correlations often play a decisive role in the spontaneous formation of these dynamical structures~\cite{Georgiev07,Goldenfeld09,Butler11,tauberLV,Dobramysl_2018}.
Consequently, various population dynamics models have been studied using stochastic lattice representations. 
Prominent examples in the literature are the Lotka--Volterra competition and coexistence of a pair of predator and prey species~\cite{Georgiev07,Dobramysl_2018,Matsuda1992StatisticalMO,Satulovsky1994StochasticLG} as well as its cyclically competing variants with additional species~\cite{Dobramysl_2018,Frachebourg1996SegregationIA,Frachebourg1996SpatialOI,He10,He11,Rulands,Szolnoki_2015,Szolnoki_2016}.

In this present work, we study the stochastic May--Leonard model \cite{May_Leonard} with three cyclically competing species on a two-dim\-ensional lattice.  
This particular model has been studied extensively; its steady-state behavior, model variations and phases in terms of the parameter regime and the extinction statistics \cite{He11,Labwavic}, effect of asymmetry \cite{avelino2021weak,Bazeia_2020}, as well as modelling its spiral pattern formation \cite{Reichenbach2008368,Reichenbach07mobilitypromotes,Serrao_2017} have been thoroughly investigated \cite{Dobramysl_2018,Szab__2007,Szolnoki_2014,Szolnoki_2020}.
Further studies addressed the robustness of biodiversity against external invasive species \cite{Blahota_2020} and the stability of ecosystems in relation to defector and cooperator alliances \cite{Szabo_2007,Szolnoki_2010}.
However, the control of these patterns and stationary states through a local, external coupling has not yet been explored in detail, especially in the (natural) regime of asymmetric interaction rates.
We show that the May--Leonard model with asymmetric predation rates is quite fragile with respect to extinction of two of the three species: 
First, the asymmetric May--Leonard variant is more likely to reach two-species extinction the stronger the asymmetry (that will be parametrized through the asymmetric factor $k$) in the predation rates is. 
In addition, for weak asymmetry in the predation rates, only a fraction of system realizations leads to a three-species coexistence state characterized by spontaneous spiral formation.

We then demonstrate through stochastic Monte Carlo lattice simulations that under the external influence of a stable May--Leonard patch, namely one that is governed by symmetric predation rates, which is spatially connected via diffusive particle spreading, one may seed conditions to ensure spiral formation in the unstable asymmetric May--Leonard system. 
This diffusive coupling of the asymmetric, vulnerable subsystem to a smaller symmetric, stable region induces planar population invasion fronts, which markedly enhances the likelihood of producing spirals for a greater range of asymmetries in the predation rates, and consequently drastically improves ecological stability. 

In the following, we first delineate the asymmetric May--Leonard model for three species on a two-dimensional lattice in section~\ref{sec:Asymmmodel}. 
In section~\ref{sec:AsymmMeanField}, we discuss the mean-field features and their limitations in capturing such a stochastic spatially extended system with a simplified mass action factorization and the resulting rate equations. 
After describing our Monte Carlo algorithm in section~\ref{MCAlgorithm}, we demonstrate spontaneous spiral structure formation in the system, and quantify the stability of the symmetric version of the stochastic May--Leonard model in section~\ref{sec:AsymmSymmetricSection}. 
Section~\ref{sec:AsymmAsymmetricSection} delves into the regime of asymmetric predation rates of the model and demarcates the conditions under which prevalent extinction of two of the three species is observed. 
In section~\ref{sec:AsymmCoupledSection}, we investigate this asymmetric model spatially coupled to a symmetric May--Leonard patch with otherwise identical parameters that seeds the asymmetric region with spirals and also enlarges the parameter range of the asymmetry factor $k$ under which spirals are observed. 
In the final section~\ref{sec:AsymmResultsSection}, we quantify more precisely the conditions under which spirals are formed, and when in contrast seeded spirals are destroyed due to the strong underlying asymmetry in the predation rates of the model.

\section{May--Leonard model}
\subsection{Model description}
\label{sec:Asymmmodel}
The cyclic May--Leonard model for three competing species \cite{May_Leonard} that we indicate with $A$, $B$, and $C$, will be simulated on a two-dimensional toroidal lattice (i.e., rectangular lattice with periodic boundary conditions) of $L_x \times L_y$ lattice cells, where $L_x$ and $L_y$ are the number of cells along the $x$ and $y$ directions. 
In order to model the finite carrying capacity of real-world ecosystems, each lattice cell is restricted to hold at most one individual (of any of the three species). 
All interactions between population members are strictly nearest-neighbor, representing the local nature of the interactions between individuals.
It follows that the allowed state for each lattice cell is $A$, $B$, $C$, or empty (denoted as $\emptyset$). 

The May--Leonard model for the three species involves their mutual competition in a cyclic pairwise predator-prey relationship,  
\begin{eqnarray}
  &A + B \to A 
  \quad &{\rm with \ rate} \ k\ \sigma \ , \nonumber \\
  &B + C \to B 
  \quad &{\rm with \ rate} \ \sigma \ ,  \nonumber  \\
  &C + A \to C 
  \quad &{\rm with \ rate} \ \sigma   ,
  \label{Rcyclic_4} 
\end{eqnarray}
with the asymmetry factor $k \in [ 0, \infty]$ parametrizing the asymmetry in the predation rate.
For $k=1$, the scheme (\ref{Rcyclic_4}) reduces to the symmetric version of the model (see Table \ref{tab:AsymmetricSpeciesNotation}).
In this work, we choose $k$ in the range $[0.1, 5]$, which is qualitatively representative of the entire space.

In addition, all species reproduce independently with a fixed rate $\mu$, which distinguishes the May--Leonard system from the cyclic Lotka--Volterra or ``rock--paper--scissors'' model \cite{Maynard82}, where predation and reproduction processes happen simultaneously and the total population number is strictly conserved,  
\begin{eqnarray}
\label{Rrepro}
  &X + \emptyset \to X + X \quad &{\rm with \ rate} \ \mu \ , 
\end{eqnarray}
where $X$ represents one of the three populations ($ X \in \{A,B,C\}$).

Finally, we add particle hopping to empty lattice sites and particle exchange reactions on the spatially extended version of the model,  
\begin{eqnarray}
\label{Rdiffuse}
  &X + \emptyset \to \emptyset + X \quad &{\rm with \ rate} \ D \ , \\
  &X + Y \to Y + X \quad &{\rm with \ rate} \ D \ .
  \label{Rswap}
\end{eqnarray}
For simplicity, we here use the same rates $D$ for both exchange and hopping reactions and $X, Y \in \{A,B,C\}$. 
These stochastic particle hopping and exchange processes on the lattice represent the general spatial mobility of species. 
In addition, we set $\sigma = \mu = 0.2$; this fixes the time scale in the system and reduces the number of independent parameters to just two, namely the asymmetry factor $k$, and the mobility rate $D$. 
We note that varying $D$ alters the fundamental length and time scale of the emerging spiral structures, whereas $k$ will parametrize their stability. 

\subsection{Mean-field analysis}
\label{sec:AsymmMeanField}
In the well-mixed limit, mean-field mass action factorization becomes applicable, and we may thus reduce the stochastic evolution of three particle types on the lattice to a set of three coupled rate equations, 
\begin{eqnarray}
\label{Asy-MF-eq}
 &&\dot{a}(t) = \mu \, a(t) - \sigma \, c(t) \, a(t) \ , \nonumber \\
 &&\dot{b}(t) = \mu \, b(t) - k \, \sigma \, a(t) \, b(t) \ , \nonumber \\
 &&\dot{c}(t) = \mu \, c(t) - \sigma \, b(t) \, c(t) \ .
\end{eqnarray}
Note that we have ignored the finite local and global carrying capacities here.
\begin{figure}
\centering
  \includegraphics[width=0.49\textwidth]{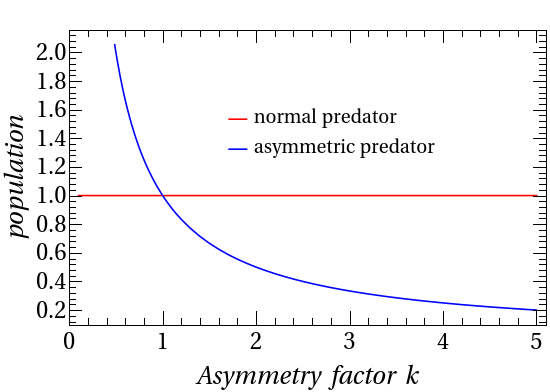}\label{fig:mfstability:a}
\caption{Steady-state mean-field population densities in units of $\frac{\mu}{\sigma}$ for the asymmetric predator (blue) and two normal predators (red) in the coexistence state. The steady state population density of the asymmetric predator is k inverse the populations of the normal predators.}
\label{fig:mfstability}   
\end{figure}
The fixed points of this mean-field system of ordinary differential equations for assumed homogeneous population densities can be computed from equating 
$\dot{a}(t)=\dot{b}(t)=\dot{c}(t)=0$ in eq.~(\ref{Asy-MF-eq}). 
The system has one unstable fixed point where all species densities vanish: it represents the absorbing state of full extinction; note that it will in contrast be the only ultimately stable state in the fully stochastic model. 
In addition, there is one three-species coexistence fixed point, $a^{*} = \mu / (k \, \sigma)$, and $b^{*} = c^{*} = \mu / \sigma$, both depicted in Fig.~\ref{fig:mfstability}. 
At this coexistence fixed point, one of the eigenvalues of the associated linear stability matrix is purely real and always negative, $\nu_{1} = - \mu $, indicating exponential relaxation to this fixed point. 
The other two eigenvalues are $\nu_{2/3} = \mu (1 \pm \sqrt{3}i) / 2$, i.e., complex conjugates with a positive real part. 
Their imaginary part is the origin of the oscillatory phenomena observed in the May--Leonard system.
As indicated by these eigenvalues, the stability of the coexistence phase is independent of the asymmetry factor $k$. 
As we shall see later, in the spatio-stochastic version of this model, the asymmetry factor $k$ plays a significant role in the oscillatory behavior of the system near coexistence.

Apart from these solutions, the mean-field approximation yields a state where two species go extinct, and only one survives,  whose density subsequently grows exponentially, i.e., the surviving species is governed by the growth equation $\dot{x}(t) = \mu \, x(t)$. 
This exponential increase in the single species mean-field density translates into a stationary state on a finite lattice, where one species fills the entire system. We note that while the stationary state corresponding to three-species coexistence is observed as a quasi-steady state for realistic times that can be simulated (or even in real-world time scales), in principle the coexistence state on a finite lattice will necessarily ultimately succumb to extinction due to large but rare fluctuations in the system.
The mean time needed to observed such large fluctuations typically scales as the exponential of the number of particles in the system and is hence irrelevant for our discussion here. 
We do not attempt to reach such time scales in the present study and shall assume that our quasi-steady coexistence state behaves for all purposes like a true stationary state of the system.

\section{Lattice simulations}
\label{sec:AsymmSimulation}
The stochastic May--Leonard model (\ref{Rcyclic_4})--(\ref{Rswap}) is simulated on a two-dimensional lattice with periodic boundary conditions in both directions. 
We initiate the Monte Carlo runs with a random configuration (fully disordered state), with parameter values that correspond to the coexistence fixed point in the mean-field version. 
The color coding of the species on the lattice is listed in Table~\ref{tab:AsymmetricSpeciesNotation}: 
\begin{table}[ht]
    \centering
    \vline
\begin{tabular}{|c|c|c|}
\hline
    A & red & asymmetric predator   \\ \hline
    B & green & prey of the asymmetric predator \\ \hline
    C & blue & predator of the asymmetric predator \\ \hline
    $\emptyset$ & white & empty lattice site \\ \hline
\end{tabular}
    \caption{Table denoting the asymmetric predator $A$ (red), its prey $B$ (green), its predator $C$ (blue), \textcolor{red}, and empty lattice sites $\emptyset$ (white). 
	All lattice images (Figs.~\ref{fig:SymmSnapshots}, \ref{fig:SymmSnapshotsVaryD}, \ref{fig:AsymmSnapshots}, \ref{fig:AsyanalysisI}, \ref{fig:CoupledSnapshots}) use this color code.}
    \label{tab:AsymmetricSpeciesNotation}
\end{table}

\subsection{Monte Carlo simulation algorithm}
\label{MCAlgorithm}
All results reported here originate from Monte Carlo simulations of the May--Leonard reaction system on a two-dimensional toroidal lattice. 
The Monte Carlo method captures the full stochastic nature of the model while maintaining a linear progression of time. 
The Monte Carlo simulation algorithm used in this paper is succinctly described as follows:
\begin{enumerate}
    \item Pick an occupied site at random from the $L_{x} \times L_{y}$ available lattice sites.
    \item Pick one of its four nearest neighbors at random with equal probability $1/4$.
    \item \begin{itemize}
    \item If that neighboring site is occupied, perform either the predation reaction (\ref{Rcyclic_4}) with probability $(k) \sigma/2$ (depending on the involved particle species) or the exchange reaction (\ref{Rswap}) with probability $D / 2$.
    \item If the neighboring site is empty, perform either the reproduction reaction (\ref{Rrepro}) with probability $\mu/2$ or the diffusion reaction (\ref{Rdiffuse}) with probability $D/2$. 
    \end{itemize}
    \item Repeat steps (1) to (3) $L_{x} \times L_{y}$ times to complete one Monte Carlo step (MCS).
    \item Repeat the above sequence until a (quasi-) stationary state is attained. 
    We assume the (quasi-) stationary state is reached when the transients in the system have died out and the fluctuations in the system have become small compared the average density in the system.
    With our system parameters, we observed all transients to decay away within $4000$ MCS. The system is then allowed to run further until $12500$ MCS.
\end{enumerate}

\subsection{Spiral structures for symmetric predation rates}
\label{sec:AsymmSymmetricSection}

\begin{figure*}
  \centering
  \subfigure[]{\includegraphics[scale=0.7]{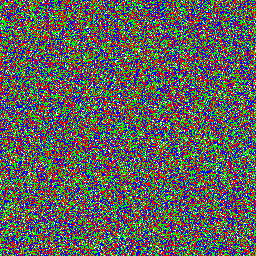}\label{}} \quad
  \subfigure[]{\includegraphics[scale=0.7]{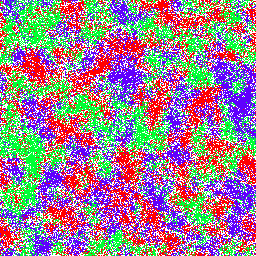}\label{}} \\
\subfigure[]
{\includegraphics[scale=0.7]{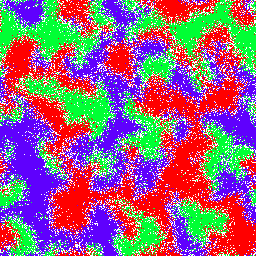}\label{}} \quad
\subfigure[]
{\includegraphics[scale=0.7]{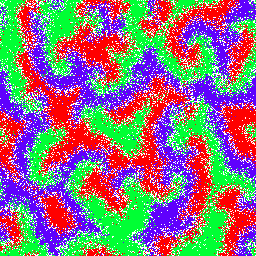}\label{}}
\caption{Snapshots of a single Monte Carlo lattice simulation with symmetric predation rate scheme, with periodic boundary conditions, at different time steps (from left to right, top to bottom) illustrating the spontaneous formation of spirals from an initial state of randomly distributed particles (a). 
Population aggregates form quickly (b), at $100$ MCS, which subsequently become more clustered (c), at $380$ MCS.
Here the interactions of populations happen only at the boundaries and the onset of waves is visible. 
These waves of three species chasing each other take the shape of (quasi-) stationary state spirals as seen in (d), at $12500$ MCS.} \label{fig:SymmSnapshots}
\end{figure*}
In the model with symmetrically chosen predation rates, $k$ is fixed to one, which renders all the species equally strong in their predation efficacy.  
Fig.~\ref{fig:SymmSnapshots} illustrates a typical stochastic May--Leonard model simulation on a two-dimensional lattice ($L_{x}=L_{y}=256$) with periodic boundary conditions with symmetric rates, $\mu=\sigma=0.2$, $k=1$, and $D=0.8$. 
(For simplicity, both $\mu$ and $\sigma$ are set to $0.2$ throughout this work).
We observe the formation of spatio-temporal spirals which represent a robust quasi-stationary dynamical state of the system starting from random initial conditions, Fig.~\ref{fig:SymmSnapshots}(a). 
At $100$ MCS (b), we notice the formation of population aggregates, followed by highly clustered single-species domains at around $380$ MCS (c). 
At this point, the system passes from the initial phase, and the evolution of the system now depends on the balance of swapping and predation reactions on the domain boundaries. 
This sets up periodic waves, which owing to the directional isotropy of the system, are manifested as spiral waves with both helicities equally likely, as seen in Fig.~\ref{fig:SymmSnapshots}(d). 
This spontaneous onset of spirals from an initial state of randomly distributed particles is typical of most symmetric versions of the May--Leonard model as long as the diameter of the emerging spirals are smaller than the underlying lattice system extension thereby avoiding extinction due to finite-size effects.

\begin{figure*}
\centering
\subfigure[]{
    \label{fig:sfig1} 
    \setcounter{subfigure}{1}
    \includegraphics[scale=0.7]{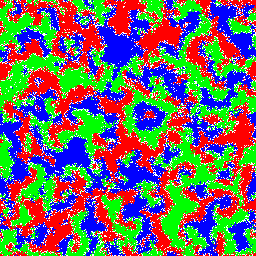}} \quad
  \subfigure[]{
    \label{fig:sfig2} 
    \setcounter{subfigure}{2}
    \includegraphics[scale=0.7]{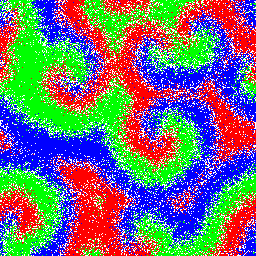}}
\caption{Simulation snapshots of the symmetric May--Leonard model with equal predation rates at both lower mobility, $D=0.1$ (a) and higher diffusivity, $D=1.5$ (b).}
\label{fig:SymmSnapshotsVaryD}
\end{figure*}

As we increase the value of $D$ (diffusion / swapping parameter), we observe an increase in the size of the spirals, see Fig.~\ref{fig:SymmSnapshotsVaryD}.
Particle aggregates set up during the initial phase are typically more extended for greater values of $D$; these later transform into spiral arms of larger thickness and wider diameter. 
We note that when the size of the spirals increases, the extent of localized oscillations also expands to eventually span the entire domain. 
At this point, further increase in the spiral size, owing to a very large diffusion / swapping parameter $D$, causes the system to approach an absorbing state of only one surviving species, and two-species extinction.
This extinction is caused by the finite size of the lattice, and in principle applies to any real-world application of the May-Leonard model as well. 

A quantitative analysis of the system is substantiated by evaluating the following quantities:
\begin{enumerate}
    \item The population densities $n_{X}$ ($ X \in \{A,B,C\}$), averaged over $100$ different Monte Carlo realizations.
    These quantities $\langle n_{X}(t) \rangle$ enable us to estimate the average time taken for the transients to die out and reach the (quasi-) stationary state of the system. 
    They measure the average density of each species versus time, which may also serve to quantify the deviation of the stochastic model from the mean-field predictions.
    \item The frequency spectrum (obtained via discrete Fourier transform) of the auto-correlation function $C_{XX}(t,s) = \langle n_{X}(t) n_{X}(s) \rangle - \langle n_{X}(t) \rangle \, \langle n_{X}(s) \rangle$, where $s$ is typically a time chosen once (quasi-) stationarity has been reached, $t \geq s$, and $n_{X} = 0, 1$ denotes the occupation number of  species $X$, averaged over all lattice sites $i$ and $100$ independent Monte Carlo simulation histories.
    In the (quasi-) stationary state, time translation invariance holds and the peaks of the auto-correlation spectrum at stationarity mark the selected characteristic frequencies. 
    The primary frequency peak in the Fourier spectrum represents the spiral oscillation frequency; if the spirals of the lattice are viewed in analogy with oscillators, then this frequency maximum must correspond to their natural frequency, assuming all oscillators to be identical.
    \item To obtain information about the typical spiral size, we compute the equal-time two-point correlation functions in the (quasi-) stationary state at different sites $r$ and $r'$,  $C_{XX}(r,r') = \langle n_{X}(r) n_{X}(r') \rangle - \langle n_{X}(r) \rangle \, \langle n_{X}(r') \rangle$, where the averages are obtained over $100$ simulation runs and taken at the same instant, once a (quasi-) stationary state has been reached. 
    From matching the decay of the correlation functions with distance $|r - r'|$ with an exponential, we can extract the correlation length $\xi_{XX}$ serving as a measure of the spiral arm thickness of species $X$: $\xi_{XX}$ is calculated as the distance from the two-point correlation function maximum to the point at which it drops to $1/e$ of its maximum value.
    \item The ensemble average of the discrete Fourier transform of the spatial image. 
    The resulting intensity peaks as function of the wave-numbers point to which characteristic wave-vectors become strongly selected in the (quasi-) stationary lattice configurations. 
\end{enumerate}

\begin{figure*}
  \centering
  \subfigure[]{\includegraphics[width=0.49\textwidth]{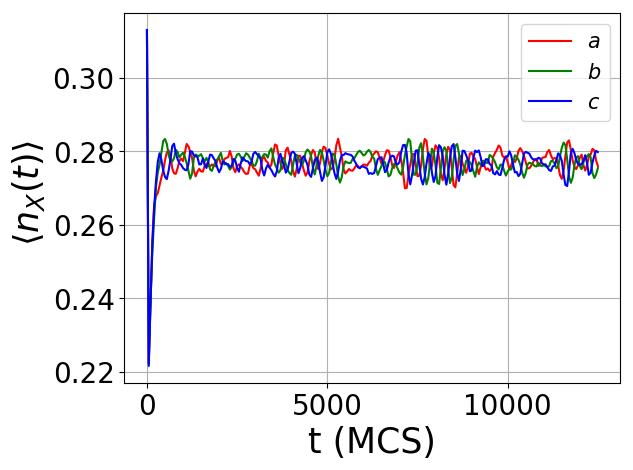}\label{fig:SyAnalysisA}}
  \subfigure[]{\includegraphics[width=0.49\textwidth]{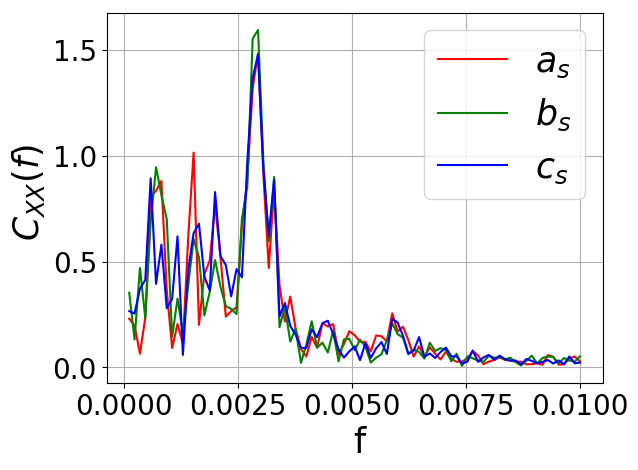}\label{fig:SyAnalysisB}}\\
\subfigure[]
{\includegraphics[width=0.49\textwidth]{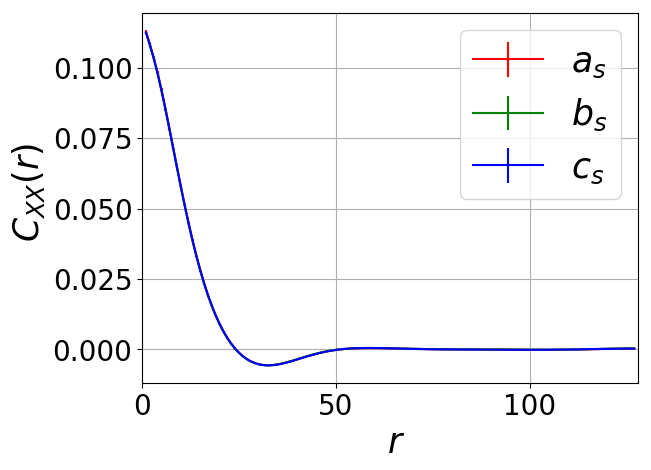}\label{fig:SyAnalysisC}}
\subfigure[]
{\includegraphics[width=0.49\textwidth]{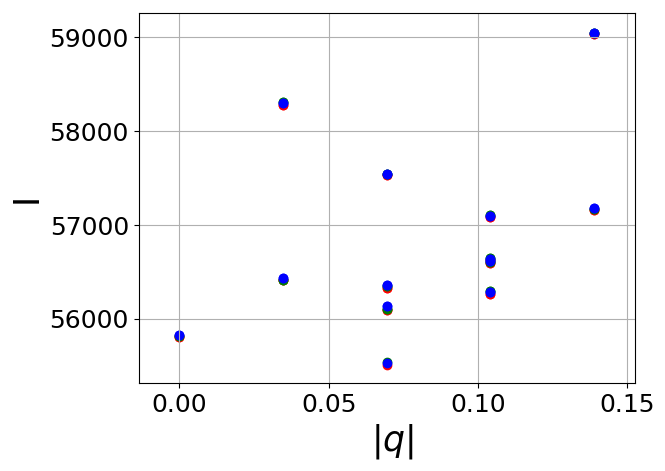}\label{fig:SyAnalysisD}}
\caption{(a) The density of the populations is stabilized after an initial transient. 
(b) The primary peak in the frequency spectrum of the auto-correlation function illustrates the typical frequencies of the oscillatory spiral waves set up in the system. 
(c) The equal-time correlation functions (indistinguishable here for all species) averaged over multiple realizations of (quasi-) stationary states indicate the characteristic size of the spirals. 
The peak intensities of the strongest wave-numbers are plotted in (d). 
The peaks at $|k|= 0.07$ correspond to the spiral wavelength of $13.15$ lattice sites. 
In this figure, $D=0.8$, $\sigma = \mu = 0.2$, and $L_{x}=L_{y}=256$.}\label{fig:Symmetricanalysis}
\end{figure*}

Although the stochastic May--Leonard model with symmetric rates is well-studied, see, e.g., Refs.~\cite{He11,Reichenbach07mobilitypromotes,Reichenbach2008368}, we recapitulate some of its features that are relevant for comparison with the asymmetric version. 
In the symmetric model, Fig.~\ref{fig:SyAnalysisA}, the transients in the density curves die out within a time period that corresponds to the onset of spiral waves in the lattice. 
Once the (quasi-) stationary spiral state is established, the instantaneous species densities merely fluctuate around the mean density.
As $D$ is increased in the symmetric model, the spiral size grows until it spans the entire system. 
At this point, finite-size effects drive the system to the extinction state.
Spectral analysis of the auto-correlation function in the (quasi-) stationary state,  Fig.~\ref{fig:SyAnalysisB}, reveals a strong spiral wave signature. 
The characteristic spiral oscillator frequency $\omega=0.00294$ MCS$^{-1}$ is obtained from the frequency peak displayed in the Fourier spectrum, c.f.~Fig.~ \ref{fig:SyAnalysisB}. 
We observe additional peaks at lower frequencies that reflect the slower motion of spiral defects due to their mutual interactions. 

In addition, the two-point equal-time correlation functions shown in Fig.~\ref{fig:SyAnalysisC} yield estimates for the thickness of the spiral arms, which in the symmetric model variant are of course identical for all three species. 
We infer the spiral arm thickness from the correlation length $\xi_{XX}$ of species $X$, which is obtained as the distance $r$ at which the two-point equal-time correlation function drops by a factor $e$. 
For $D=0.8$, $\xi_{XX} \approx 13.15$ in lattice units. As the value of $D$ is raised, the correlation length $\xi_{XX}$ increases, commensurate with the larger diffusivity and hence greater spiral size. 
However, once the spiral extent reaches the system size, $\xi_{XX}$ saturates to a fixed value. 
Finally, we report the wave-numbers of the averaged signal of the images at (quasi-) stationarity in Fig.~ \ref{fig:SyAnalysisD}; the weights of the selected wave-numbers are reflected in the associated intensity peaks. 
The peaks at $|k|=0.07$ correspond to a wavelength of $\sim 13$ lattice sites, which matches the correlation length obtained from Fig.~\ref{fig:SyAnalysisC}.     

\subsection{Asymmetric rates: extinction vs. spiral structures}
\label{sec:AsymmAsymmetricSection}
%

\begin{figure*}
    \centering
    \subfigure[]{\includegraphics[scale=0.7]{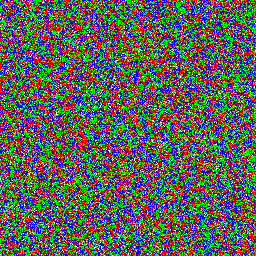}} \qquad
    \subfigure[]{\includegraphics[scale=0.7]{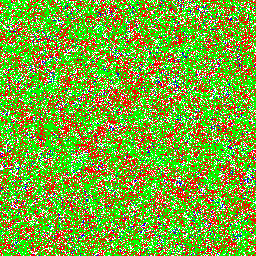}} \\ 
    \subfigure[]{\includegraphics[scale=0.7]{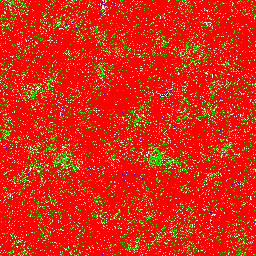}} \qquad
    \subfigure[]{\includegraphics[scale=0.7]{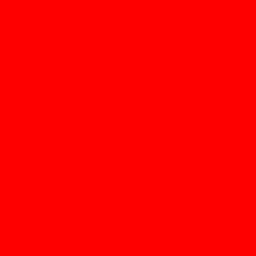}} \\
    \caption{Snapshots of a single Monte Carlo lattice simulation (from left to right, top to bottom) of the May--Leonard model with asymmetric rates ($D = 0.8$ and $k = 0.5$) leading to the extinction of two species. 
    With asymmetrically set rates, initial cluster aggregates of the three species (a) are replaced by initial dominance of a single (green) species (b), at $100$ MCS), namely either the prey of the weakest predator (for $k < 1$), or the strongest predator ($k > 1$).
    Abundance of this species then causes its predator population (red) to rise (c), at $120$ MCS, and ultimately dominate the system. 
    This in turn enables its predator population to increase, etc.: one observes a heteroclinic cycle until one of the species goes extinct owing to a stochastic fluctuation event, after which the cycle becomes broken and only a single species (here, red) survives (d).} \label{fig:AsymmSnapshots}
\end{figure*}
\begin{figure*}
  \centering
  \subfigure[]{\includegraphics[scale=0.7]{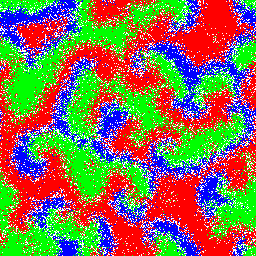}\label{fig:AsyAnalysisIA}} \qquad
  \subfigure[]{\includegraphics[scale=0.7]{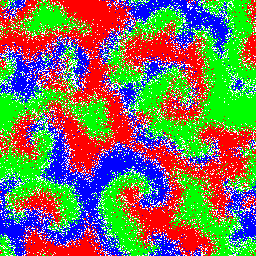}\label{fig:AsyAnalysisIB}}\\
\subfigure[]
{\includegraphics[height=5cm,width=0.44\textwidth]{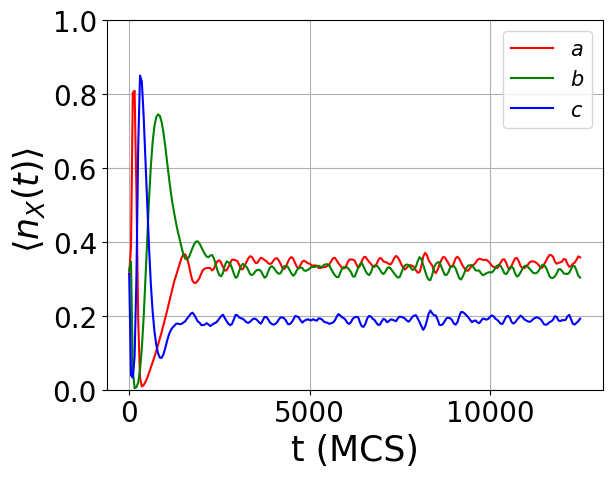}\label{fig:AsyAnalysisIC}}
\subfigure[]
{\includegraphics[height=5cm,width=0.44\textwidth]{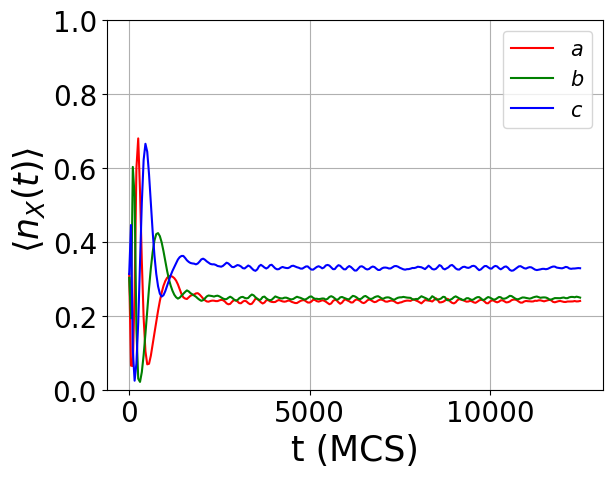}\label{fig:AsyAnalysisID}}\\
\subfigure[]
{\includegraphics[height=5cm,width=0.44\textwidth]{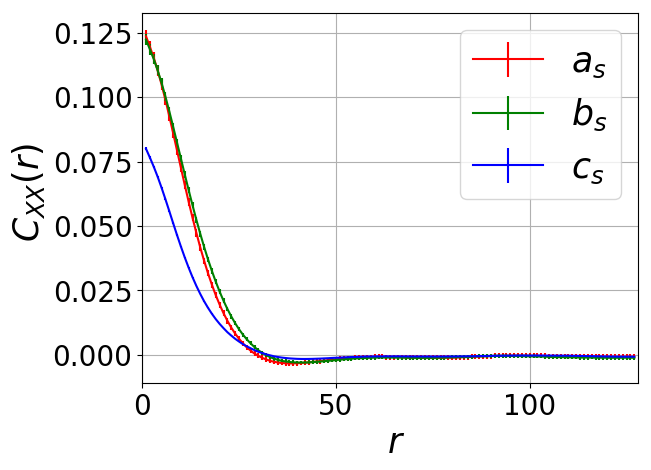}\label{fig:AsyAnalysisIE}}
\subfigure[]
{\includegraphics[height=5cm,width=0.44\textwidth]{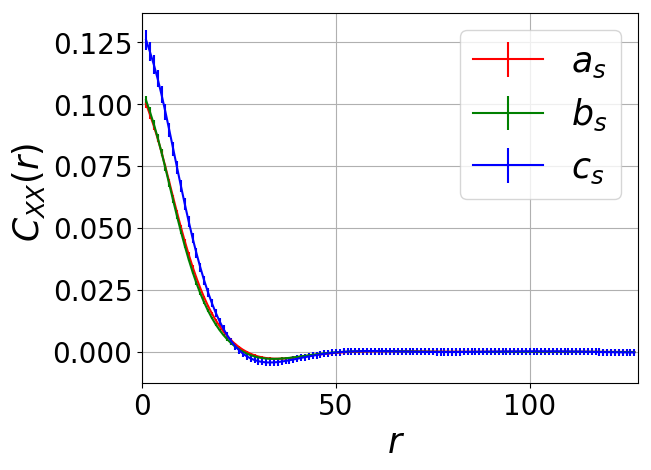}\label{fig:AsyAnalysisIF}}
\caption{Snapshots of typical simulations for the (quasi-) stationary state in the asymmetric May--Leonard model for asymmetry parameters (a, c, e) $k = 0.5 < 1$ and (b, d, f) $k = 1.5 > 1$. 
The density of the predator of the asymmetric predator (blue) is anomalous in both cases. 
(c) For $k < 1$, the density of the blue species is lower than both the red-green combination in the (quasi-) stationary state. 
(d) For $k > 1$, the blue predator assumes a higher (quasi-) stationary state density. 
The equal-time correlation functions at (quasi-) stationarity show similar trends: 
(e) For $k < 1$, the peak of the correlation function is lower for the blue population. 
(f) For $k>1$, the blue predator correlations display a higher peak value.} \label{fig:AsyanalysisI}
\end{figure*}
\begin{figure*}
  \centering
  \subfigure[]{\includegraphics[width=0.49\textwidth]{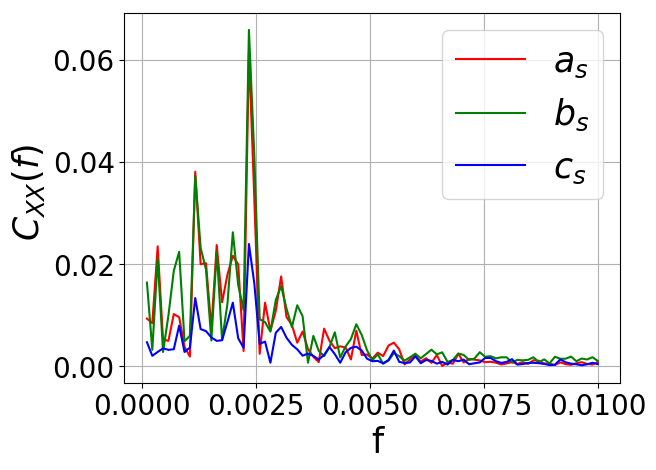}\label{fig:AsyAnalysisIIA}}
  \subfigure[]{\includegraphics[width=0.49\textwidth]{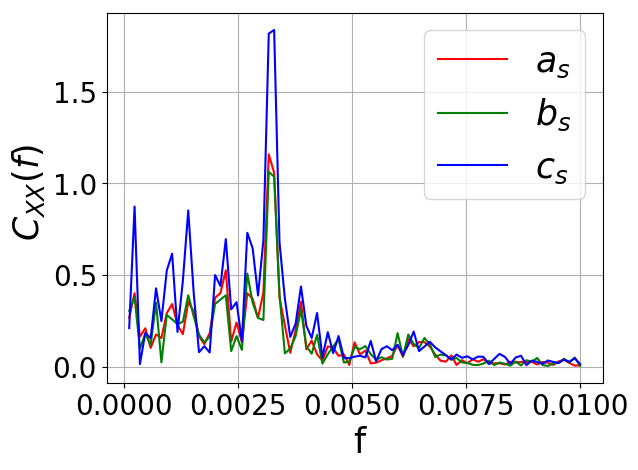}\label{fig:AsyAnalysisIIB}}\\
\subfigure[]
{\includegraphics[width=0.49\textwidth]{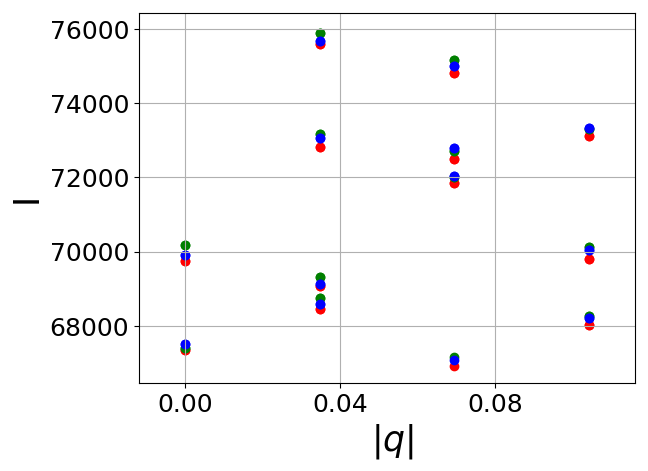}\label{fig:AsyAnalysisIIC}}
\subfigure[]
{\includegraphics[width=0.49\textwidth]{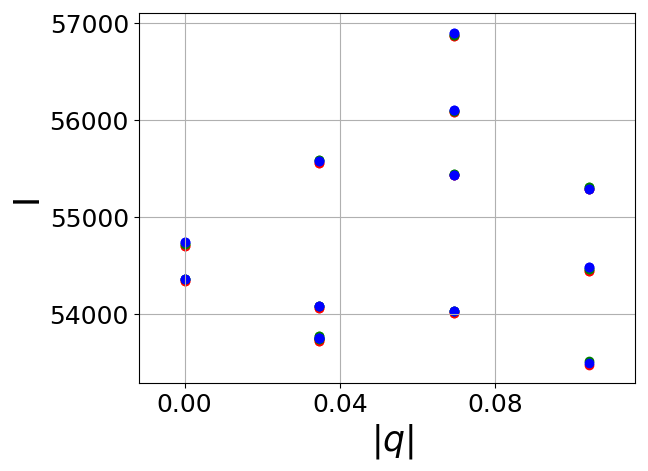}\label{fig:AsyAnalysisIID}}
\caption{The frequency spectrum of the auto-correlation function in these symmetric May--Leonard model variants shows that for (a,c) $k = 0 .5 < 1$ the blue species (predator of the asymmetric predator) displays a lower peak intensity at the same frequency  than the other two species (red and green). 
(b,d) For $k = 1.5 > 1$, this situation is reversed. 
We observe that the selected wavelengths peak at different frequencies.} \label{fig:AsyanalysisII}
\end{figure*}
Asymmetric stochastic May--Leonard model variants are designed to reproduce more realistic cyclic predator prey dynamics where the predatory propensities of each species differ.
Simulations of the May--Leonard model for asymmetric predation rates generally yield greater proclivity towards the absorbing state of a single surviving population ultimately dominating the entire lattice. 
However, for weakly asymmetric predation rates and intermediate diffusivities, one can still observe the presence of spirals for a subset of stochastic realizations (independent Monte Carlo simulation runs). 
The precursors to stable spatio-temporal spiral patterns are the initial globular clusters of all three species; the size of these clusters is typically much smaller than the system size. 
These initial clusters drive the individuals at the interfaces to chase each other and thus set up spiral waves in the system. 
The presence of clusters in the initial time depends crucially on two system parameters, namely the diffusion / swapping rate $D$ and the asymmetric predation factor $k$.
(We remark that extinction probabilities also depend on the initial state of the lattice; yet this work exclusively discusses disordered, random initial configurations.) 
As in the symmetric case, for large $D$ the system is deprived of localized clusters and the populations oscillate on a global level, akin to the zero-dimensional non-spatial limit. 
On the other hand, for extreme values of the asymmetric predation factor, see Fig.~\ref{fig:AsymmSnapshots}, one species  dominates the system in the early-time regime precluding the emergence of localized clusters. 
This abundance of the dominant species at a global level then enhances its predator population which in turn boosts its own predator in the subsequent time evolution. 
Thus the system oscillates from one dominant species to another until one of the species goes completely extinct, resembling the heteroclinic cycles of the corresponding deterministic system of coupled mean-field rate equations~\cite{Labwavic,He11}. 
The final extinction event however constitutes a purely stochastic phenomenon and cannot be reproduced with a mere deterministic treatment of the model. 

For weakly asymmetric predation rates, spiral pattern formation appears in a similar manner as for the purely symmetric May--Leonard system. 
Even in this scenario, finite systems either reach the absorbing two-species extinction state or the (quasi-) stationary state of three-species coexistence; the outcome is probabilistic in nature and the likelihood of extinction depends on various system parameters. 
In the following analysis, we restrict ourselves to those realizations that lead to a (quasi-) steady coexistence state, whose asymptotic features depend highly on the asymmetry factor $k$. 
For $k < 1$, c.f.~Figs.~\ref{fig:AsyAnalysisIA}, \ref{fig:AsyAnalysisIC}, the density of the predator of the asymmetric species (indicated in blue, with characteristic correlation length $\xi = 13.50$) saturates at a lower value compared with the asymmetric species (red, $\xi = 14.5$) and its prey (green, $\xi = 15.75$). 
For $k > 1$, Fig.~\ref{fig:AsyAnalysisIB}, \ref{fig:AsyAnalysisID}, the converse happens:
The density of the blue species ($\xi = 13.59$) saturates at a higher value than the asymmetric predator (red, $\xi = 12.86$) and its prey (green, $\xi = 12.27$). 
This is reflected in the correlation lengths $\xi_{XX}$ computed from the equal-time correlation function at (quasi-) stationarity, plotted in Figs.~\ref{fig:AsyAnalysisIE} and \ref{fig:AsyAnalysisIF}. 
In general, for $k < 1$, the spiral arm thickness (proportional to the correlation length) of the blue species is smaller than for the red and green populations. 
For $k > 1$, the situation is reversed and the blue population displays thicker spiral arms.
We note that this is in direct contrast to the mean-field prediction for the hyperbolic unstable coexistence state of the asymmetric predator red having the greater (lesser) density for $k < 1\ (k > 1)$. 
Thus we observe here a purely stochastic coexistence (quasi-) stationary state which is comparatively stable (barring rare large deviations driving the system to two-species extinction). 

In addition, we obtain a similar contrast in the frequency peak of the auto-correlation spectrum, Figs.~\ref{fig:AsyAnalysisIIA}, \ref{fig:AsyAnalysisIIB}. 
For $k < 1$, the frequency peaks of the predator of the asymmetric predator (blue) display lower intensities than for the asymmetric predator (red) and its prey (green). 
For $k > 1$, the blue populations record a higher intensity. 
The frequencies measured at these peaks are similar across all species for a given value of $k$, e.g., for $k=0.5$, $\omega = 0.00235$ MCS$^{-1}$, whereas for $k=1.5$, $\omega = 0.00317$ MCS$^{-1}$.

\begin{figure*}
    \centering
    \subfigure[]{\includegraphics[width=0.49\textwidth]{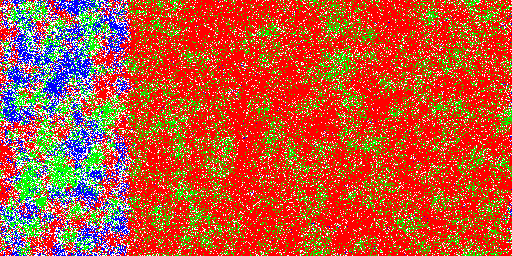}} \quad
    \subfigure[]{\includegraphics[width=0.49\textwidth]{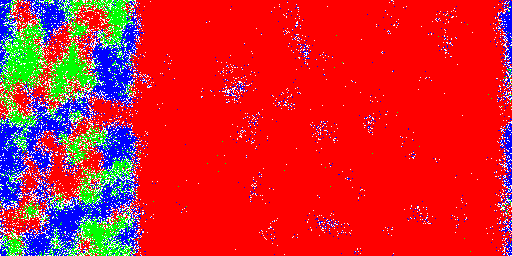}}\\
    \subfigure[]{\includegraphics[width=0.49\textwidth]{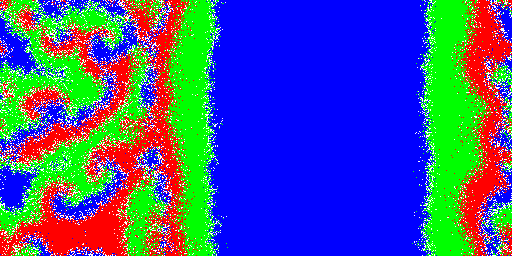}} \quad
    \subfigure[]{\includegraphics[width=0.49\textwidth]{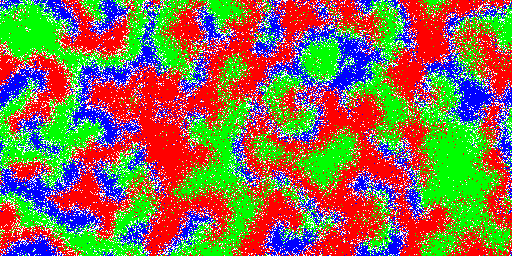}}\\
    \caption{Snapshots of typical Monte Carlo simulation (left to right, top to bottom) of a diffusively coupled symmetric May--Leonard patch ($L_y=256,L_x=128$) and asymmetric subsystem ($L_y=256,L_x=384$), thrice as large, with $D=0.8$ and $k=0.5$.
    At 100 MCS (a), the asymmetric region already shows dominance of one species (red), indicating the subsystem's instability towards extinction of two species. 
    The snapshot at time 220 MCS (b) shows that cluster formation is pronounced in the smaller stable symmetric patch in contrast with the cyclic dominance emerging in the asymmetric region. 
    We observe that planar waves emanating from the symmetric region begin to enter the asymmetric subsystem. 
    These waves enter the asymmetric region from both sides as seen at 940 MCS (c). 
    At (quasi-) stationarity (d), stable spiral structures are seeded in the asymmetric region, and persist throughout the simulation run.}
    \label{fig:CoupledSnapshots}
\end{figure*}
\section{Spatially inhomogeneous coupled May--Leonard system}
\label{sec:AsymmCoupledSection}

In view of our goal to potentially control the spiral structures observed in the May--Leonard model without globally modifying the fundamental underlying stochastic reaction rates, we explore the effect of coupling a spatial May--Leonard system with asymmetric predation rates to another one with symmetric rates.
Although the symmetric May-Leonard model favors ecological diversity and spiral patterns, cyclic models in the real world that display symmetric rates are of course extremely unlikely; yet the more viable models with asymmetric predation rates show a much greater propensity for species extinction and spatial homogeneity, suppressing species diversity.
Our motivation is therefore to control the asymmetric regime of the May--Leonard model towards a stable coexistence of all three species through maintaining robust conditions for spiral creation and survival. 
It turns out that by diffusively coupling the asymmetric May--Leonard model to a smaller stable region with symmetric reaction rates, one may not only create conditions for evading extinction in the asymmetric subsystem, but also generate spiral patterns in that region for all realizations, even when it is governed by strongly asymmetric predation rates. 

\begin{figure*}
  \centering
  \subfigure[]{\includegraphics[height=5cm,width=0.40\textwidth]{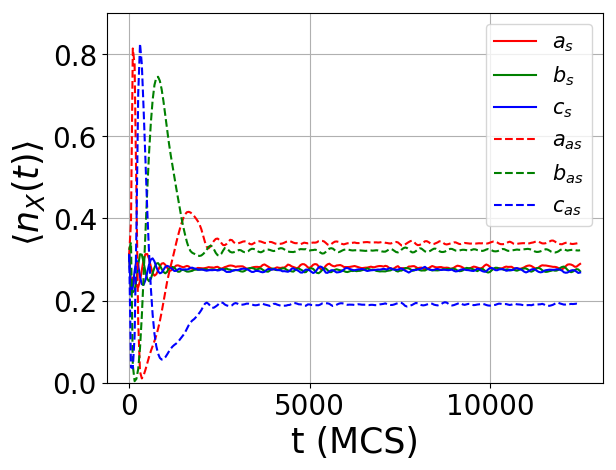}\label{fig:CoupAnalysisA}}
  \subfigure[]{\includegraphics[height=5cm,width=0.40\textwidth]{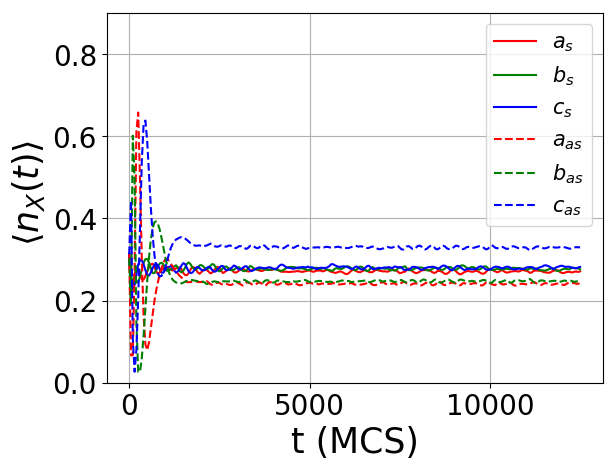}\label{fig:CoupAnalysisB}}\\
\subfigure[]
{\includegraphics[height=5cm,width=0.40\textwidth]{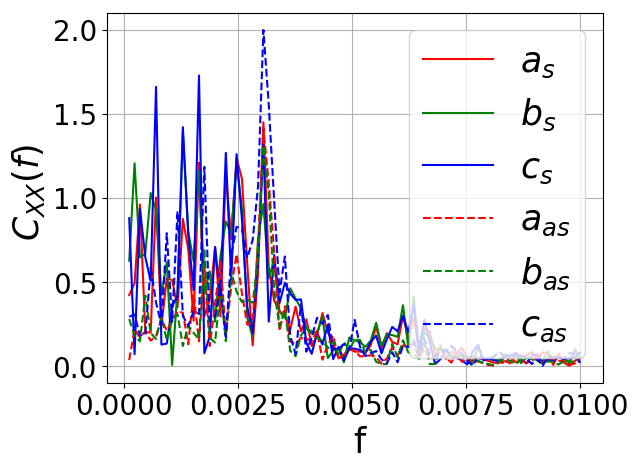}\label{fig:CoupAnalysisC}}
\subfigure[]
{\includegraphics[height=5cm,width=0.40\textwidth]{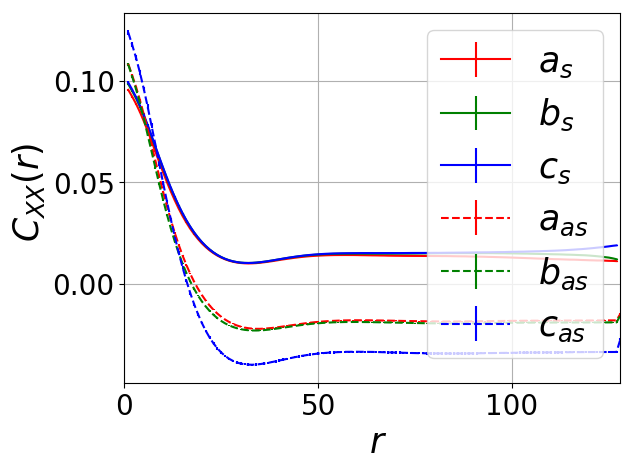}\label{fig:CoupAnalysisD}}\\
\subfigure[]
{\includegraphics[height=5cm,width=0.40\textwidth]{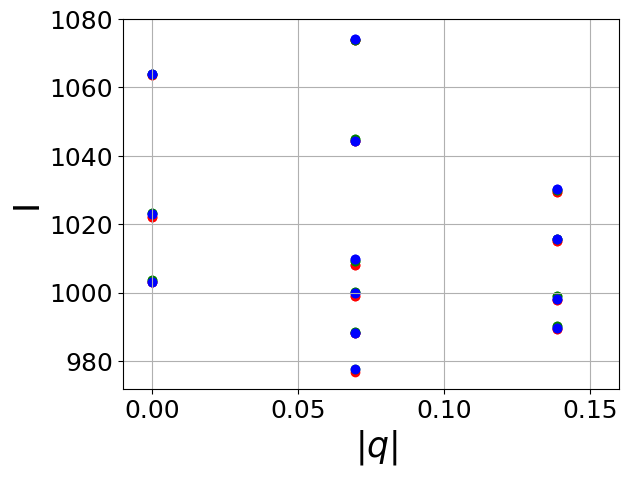}\label{fig:CoupAnalysisE}}
\subfigure[]
{\includegraphics[height=5cm,width=0.40\textwidth]{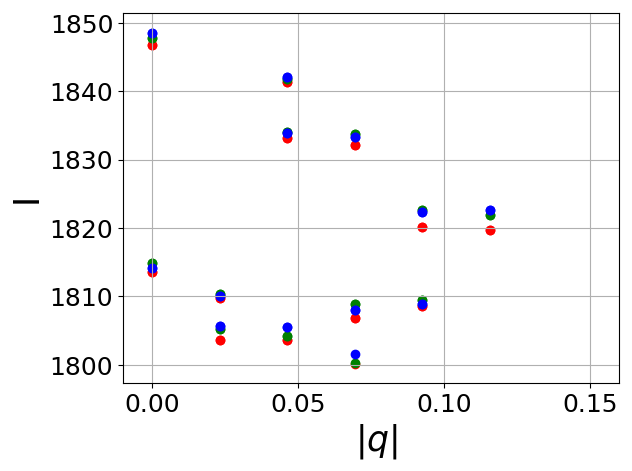}\label{fig:CoupAnalysisF}}
\caption{Coupled May--Leonard patches of a stable subsystem with symmetric rates and an unstable subsystem with asymmetric predation rates for both $k < 1$ (left) and $k > 1$ (right), with representative rates $k = 0.5$ (a) and $k = 1.5$ (b). 
(a, b): population density plots; the asymmetric subsystem shows the same trends as the purely asymmetric system for those realizations which lead to spiral structures. 
Panel (c) shows additional low frequency peaks for $k = 1.5$ (at $0.0008$ MCS$^{-1}$ and $0.0017$ MCS$^{-1}$) in the auto-correlation signal, which correspond to invasive planar waves as well as moving spirals. 
These signals are stronger than in the purely asymmetric case. 
Panel (d) depicts the correlation function for the asymmetric model with $k = 1.5$. 
The two-point correlation function saturates at a much lower value. 
Figs.~(e,f) show the wavenumber selection in the symmetric and asymmetric regions, respectively. 
The wavenumbers are highly selected only for a narrow range in the symmetric region; for the asymmetric region they are in contrast broadly distributed, which is due to the invading planar wavefronts. Note : The intensities of the peaks are different for the two subsystems due to different sized patches, although the wavenumbers on the x-axis convey the most relevant physics.}
\label{fig:CoupAnalysis}
\end{figure*}
The symmetric subsystem of the May--Leonard model is implemented on a patch of the full two-dimensional square lattice (which is subject to periodic boundary conditions), spanning the whole dimension along the $y$ direction and just $1/4$ of the total system size in the $x$ direction. 
The asymmetric region $(L_{x} = 384, L_{y} = 256)$ is thus set up to have thrice the area of the symmetric model $(L_{x} = 128, L_{y} = 256)$, in order to allow for the spontaneous emergence of spiral structures in the symmetric region, while maintaining the two symmetric-asymmetric subsystem boundaries at sufficient distance to minimize immediate communication between the interfaces. 
The parameters of the symmetric subsystem are chosen identical to that of the asymmetric model, except that in the asymmetric model $k$ differs from $1$. 
The other parameters $\mu = \sigma = 0.2$ and $0.1 \leq D \leq 1.8$ are set to facilitate the formation of  robust spirals in the symmetric subsystem (by keeping its control parameters away from the two-species extinction caused by large diffusivities in small systems). 

The interplay of the two subsystems and the induction of a coexistence (quasi-) stationary state in the asymmetric region is demonstrated in the time sequence in Fig.~\ref{fig:CoupledSnapshots}. 
The contrast in the initial behavior of the two subsystems is apparent in the top row, (a, b). 
The initial aggregate seeds that are essential to the subsequent spiral formation, crucially absent in the asymmetric region, are effectively compensated in the larger unstable patch through a periodic and cyclic migration of the three populations into the asymmetric region (c). 
This cyclic chasing mechanism of the three species at the boundary of the predator-prey interface that acts a precursor to the spiral state is vital to the formation of spirals: 
The symmetric subsystem acts as a continuous seeding generator to the asymmetric region. 
Once the periodic wave fronts meet each other in the asymmetric region, they establish robust spiral structures (d). 
Even for weak asymmetry, in the purely asymmetric model only a finite set of realizations are able to replicate the initial aggregate stage and hence reproduce stable spiral formation. 
However, for the diffusively coupled system, the conditions required for spiral formation are always reached in the asymmetric subsystem owing to seeding of cyclic periodic wave fronts from the smaller symmetric region.
Thus even in the strongly asymmetric limit, stable spirals are created in the larger asymmetric region of the symmetric-asymmetric coupled system. 

It is important to note that the spiral structures in the asymmetric subsystem are locally determined by the asymmetric predation rates. 
That is, once the spirals are seeded, their spatial structure and temporal evolution is bound to the local rules of the asymmetric model. 
Each spiral formed has three spiral bands (arms) of red-green-blue populations cyclically chasing each other with the spiral center acting as a source or defect. 
The robustness of the spiral once produced depends on the stability of this spiral defect center. 
As spiral wave bands travel outward, the discrepancy in their predation rates relative to the swapping rate produces a variability in the velocity of the wave boundaries. 
In particular, the interface between the asymmetric predator and its prey (red-green) will have a different speed than both other (blue-red and green-blue) interfaces. 
For $k < 1$ $(k > 1)$, the red-green interface is slower (faster) than the green-blue and blue-red interfaces.
This speed difference dynamically alters the thickness of the three wave fronts as the spiral arms propagate away from the central defect. 
For $k < 1$, the asymmetric predator (red) arm becomes thinner in time until its thickness goes to zero. 
For $k < 1$, the asymmetric arm becomes thicker at the expense of the blue and green arms until they both vanish. 
If the speed is fast enough to overcome even one of the spiral arms before it leaves the spiral center and migrates to the outer region of the spiral vicinity, the spiral defect is destroyed. 
Hence, spirals in the asymmetric model can be observed when coupled to a symmetric subsystem as long as the interface boundaries are kept apart to avoid destabilizing finite-size effects. 
We also point out that although a spiral defect might be annihilated due to the underlying instability of the model, new spirals are always created in its place due to the continuous injection of cyclically periodic wave fronts emanating from the symmetric subsystem. 

Thus far we have seen that the coupling to a symmetric subsystem markedly increases the stability of the larger region with asymmetric reaction rates, enlarging the range of the asymmetry factor $k$ under which spiral structures are reliably produced, thus promoting biodiversity in the system. 
Once a spiral defect is seeded by the periodic waves from the symmetric region, the thus stabilized spiral structures will evolve in the asymmetric region according to the local rules of the asymmetric May--Leonard model. 
These observations can be verified quantitatively by examining the density curves in Figs.~\ref{fig:CoupAnalysisA}, \ref{fig:CoupAnalysisB}. 
The (quasi-) stationary densities are different for the asymmetric model in comparison to the symmetric subsystem. 
Moreover, for $k < 1\ (k > 1)$, both species linked with the asymmetric rate (red and green), have a higher (lower) (quasi-) stationary state density than the blue species. 
This is very similar to observations for those (few) realizations that produce spirals in the purely asymmetric May-Leonard model.
The (quasi-) stationary state densities for the asymmetric subsystem in the coupled model are close to those measured for the realizations that produce spirals for a purely asymmetric system in the limit of weakly asymmetric rates, except that the density fluctuations in the coupled system are damped by the random phases of plane waves entering from the symmetric patch into the asymmetric region, with the phase averaging effectively reducing the mean amplitude of the spiral waves in the asymmetric subsystem. 

We observe similar trends in the frequency spectrum of the auto-correlation function of the asymmetric subsystem, Fig.~\ref{fig:CoupAnalysisC}. 
The primary peaks at the characteristic oscillation frequencies are accompanied by additional low-frequency peaks pointing to the influx of periodic waves into the asymmetric region.
Comparative trends in the two-point correlation function of the symmetric and asymmetric subsystems in the coupled case are similar to their isolated counterparts, see Fig.~\ref{fig:CoupAnalysisD}. 
In addition to the extra frequencies detected in the asymmetric region, one observes a broader selection of wavelengths in the asymmetric subsystem, as apparent in Fig.~\ref{fig:CoupAnalysisF}. 
The amplitudes for these wavelengths are not as intense as the highly selective wavelengths observed in the symmetric subsystem, Fig.~\ref{fig:CoupAnalysisE}. 
This confirms the lopsided effect that the symmetric region has on the asymmetric subsystem as opposed to the minute reverse influence of the larger asymmetric patch on the stable symmetric region.  
\begin{figure*}
  \centering
  \subfigure[]{\includegraphics[height=4cm,width=0.40\textwidth]{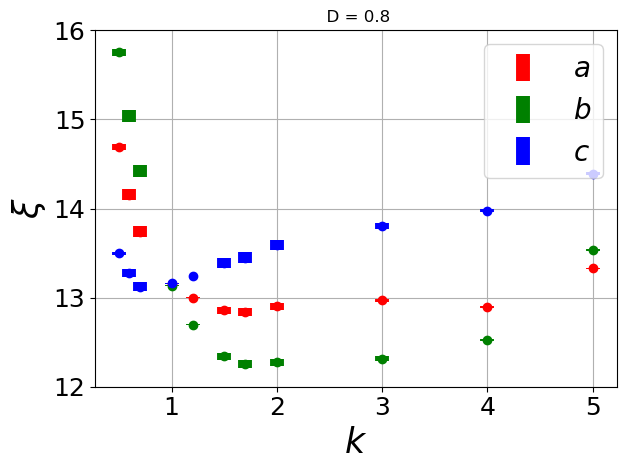}\label{fig:coupledresultsA}}
  \subfigure[]{\includegraphics[height=4cm,width=0.40\textwidth]{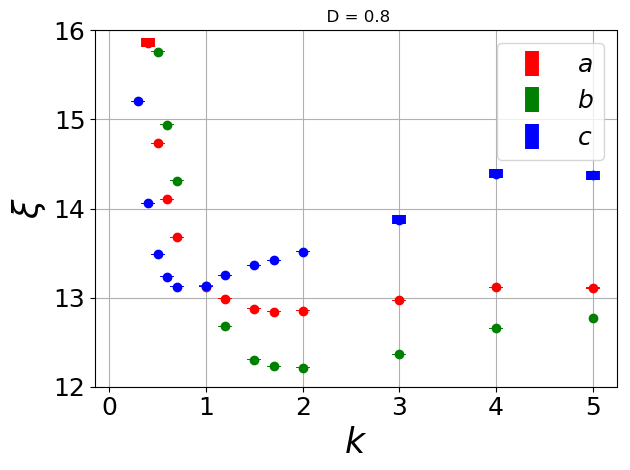}\label{fig:coupledresultsB}}\\
\subfigure[]
{\includegraphics[height=4cm,width=0.40\textwidth]{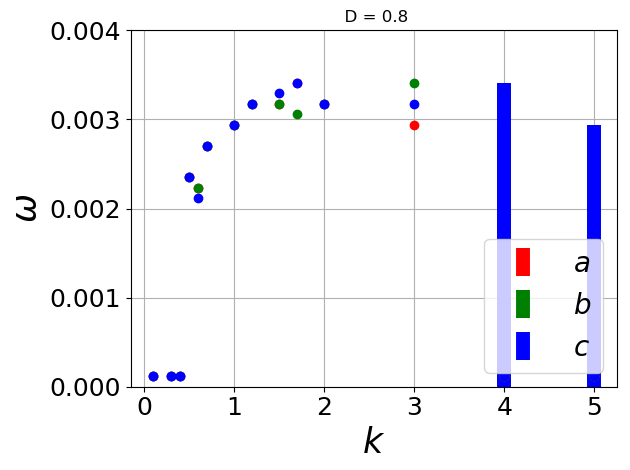}\label{fig:coupledresultsC}}
\subfigure[]
{\includegraphics[height=4cm,width=0.40\textwidth]{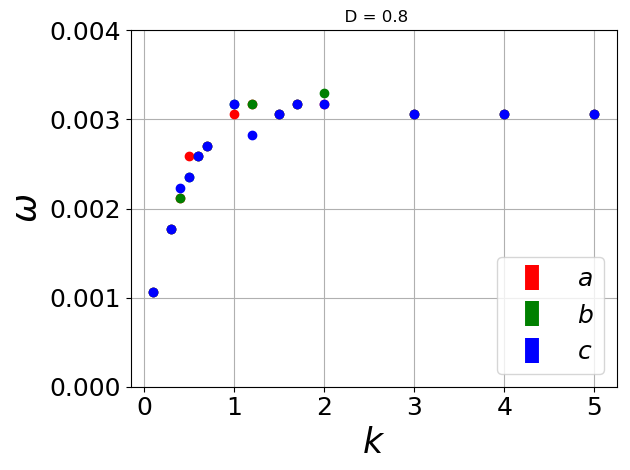}\label{fig:coupledresultsD}}\\
\subfigure[]
{\includegraphics[height=4cm,width=0.40\textwidth]{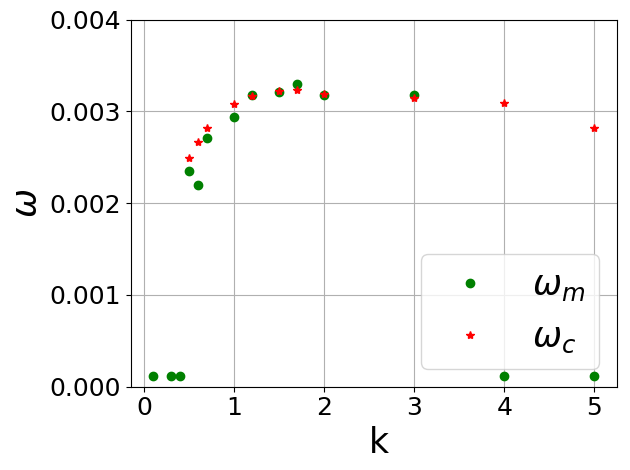}\label{fig:coupledresultsE}}
\subfigure[]
{\includegraphics[height=4cm,width=0.40\textwidth]{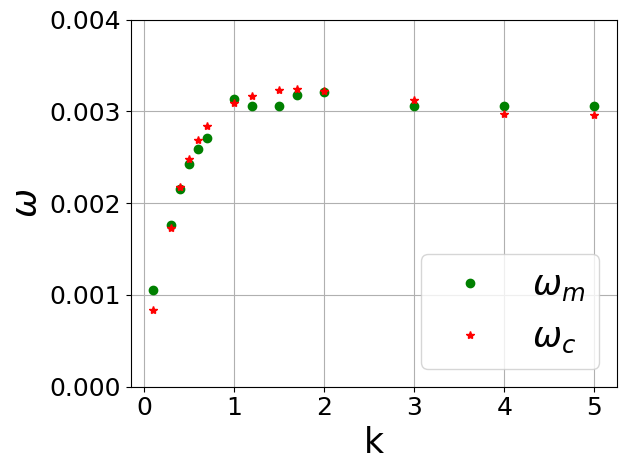}\label{fig:coupledresultsF}}
\caption{Panels (a-f) illustrate the enhanced stabilization of spiral structures in the coupled symmetric/asymmetric May--Leonard lattice for a representative diffusion rate $D = 0.8$:  
(a,b) show the comparison for the correlation lengths computed in the purely asymmetric system and in the asymmetric region of the inhomogeneous coupled lattice. 
The correlation lengths are reduced as $k$ approaches unity and display opposite trends in deviations to either direction about the symmetric point $k = 1$. 
(c,d) plot the oscillation frequencies for both the purely asymmetric system and the asymmetric region in the coupled model. 
The large error bars for $k = 4,5$ are due to poor statistics on account of extinctions observed in a large number of realizations, and spiral formation occurring in contrast only in a few realizations.
The coupled region tends toward higher stability as the oscillation frequencies saturate. 
We see similar trends of increased spiral robustness for extreme values of asymmetric rates for all variations in $D$. 
In (e,f), we display the criterion for spiral sustainability for the pure asymmetric model and the asymmetric subsystem in the coupled case, respectively; $\omega_m$ denotes the computed / measured oscillation frequency of the spiral obtained from the auto-correlation spectrum, while $\omega_{c} = 2 D / 3 \langle \xi \rangle^{2}$ is the critical inverse time scale. 
Collapse of these two sets of data points implies the stabilization of spirals for that value of $k$. 
In (e), we see that spirals are only sustained for the asymmetry interval $0.5 \leq k \leq 3$ (for a fraction of realizations).
(f) shows significant data collapse between $\omega_{m}$ and $\omega_{c}$ which corroborates the spiral structure stabilization in the coupled system.}
\label{fig:coupledresults}
\end{figure*}

We have thus demonstrated that in our spatially inhomogeneous system wherein a (smaller) symmetric patch of the cyclic May--Leonard model is coupled to a (larger) region with asymmetric rates through diffusive particle exchange, the conditions for spiral formation in the asymmetric region are markedly improved, as planar invasion fronts enter this unstable regime which is prone to two-species extinction.
In addition, the range of the asymmetry factors under which such spatio-temporal patterns can be observed is significantly broadened. 
Since the formation of these spiral patterns of subsequent predation fronts represent the basic mechanism to promote species coexistence in the cyclic predator-prey May--Leonard system, ecodiversity thus becomes considerably enhanced in what in isolation would have constituted a fragile ecosystem.

We finally provide a quantitative description for this remarkable extension of species diversity robustness in the asymmetric May--Leonard model owing to its diffusive coupling to a stable symmetric patch. 
To this end, we have computed the correlation lengths $\xi_{XX}$ of the asymmetric patches from the equal-time two-point correlation function. 
As seen in Figs.~\ref{fig:coupledresultsA} and \ref{fig:coupledresultsB}, the correlation lengths of course converge as the symmetric point $k = 1$ is approached from the regime with $k < 1$, but then diverge again in the opposite direction for $k > 1$.
Although the trends in the correlation lengths are similar in both settings, the regime for which we see spirals is clearly greater for a coupled inhomogeneous system than in a single uniform patch. 
In addition, only a small percentage of realizations render coexistence (quasi-) stationary states in the pure system, whereas in the coupled system such spiral structures are ubiquitously present in all realizations. 
The characteristic oscillation frequency for the spiral structures is computed from the frequency peak of the auto-correlation spectrum, Figs.~ \ref{fig:coupledresultsC}, \ref{fig:coupledresultsD}. 
The oscillation frequencies for the three species turn out less varied as compared to the correlation lengths; they increase as the asymmetry factor $k$ is raised from small values $\approx 0.1$, but saturate at about $k = 2$. 
Due to the stabilizing presence of the symmetric patch, the coupled system displays a greater robustness to variations of the asymmetric predation rates.

Although the spiral patterns are seeded in the asymmetric region of the coupled system, the evolution of the spirals is subject to the local parameters of the asymmetric subsystem. 
For extreme values of $k$, seeded spiral defects are annihilated because of the gradual thinning of one of the spiral arms.
In order to maintain viable spiral structures, this spiral arm rotation frequency $\omega$ must be greater than the characteristic inverse time scale of destruction of the spiral effected by diffusive particle transport through the spiral arms, which should hence be proportional to the inverse square of the average correlation length for all three species, $\omega > \omega_c \sim D / \langle \xi^{2} \rangle$. 
These comparisons are depicted in Figs.~\ref{fig:coupledresultsE} and \ref{fig:coupledresultsF} for $D = 0.8$ for the solitary and coupled models, respectively. 
Figure~\ref{fig:coupledresultsE} shows that in the isolated asymmetric system, spirals are only sustained for $0.5 \leq k \leq 3$.
For extreme values of $k$, there is an extremal value of $\xi$ for the surviving population (the other two species go extinct) corresponding to the minimal values of $\omega_{m}$.
A general sense of the extent of robustness of the spatio-temporal spiral structures may thus be estimated from the above criterion.
We note that it is important to distinguish the spatio-temporal stability of individual spirals in the asymmetric subsystem with the overall ecodiversity of the populations as inferred from the average density of the three species in the entire asymmetric region.
For extreme asymmetry factor values, the spiral defects become quickly annihilated in the asymmetric region of the coupled model.
Hence, if the plane waves invading from the symmetric region would not provide a continuous influx of individuals from all three cyclically competing species, the spirals in the asymmetric region would die out. 
It is the steady diverse population supply that constantly creates spiral structures in the asymmetric region, even for extreme values of $k$. 
Hence, these migrations play a decisive role in three-species coexistence and thus ecodiversity maintenance of the asymmetric spatially extended May--Leonard model.

\section{Conclusions and outlook}
\label{sec:AsymmResultsSection}

In this work, we shed light on the important problem of maintaining ecodiversity in a paradigmatic simple system, namely the three-species May--Leonard model, which encapsulates cyclic predation motives that are encountered in nature. 
Unequal predation rates in the May--Leonard model of course represent the more realistic scenario, where typically at least one predator has a different predation propensity in comparison to the other species.
However, such asymmetric May--Leonard systems are highly likely to reach an absorbing state corresponding to two-species extinction, with the extinction likelihood growing with the rate asymmetry.

Yet we demonstrate here that one can drastically mitigate this extinction probability by diffusively coupling the system to a smaller stable patch governed by the symmetric May--Leonard model variant, which markedly enhances the robustness of the combined model with respect to three-species coexistence.
Our Monte Carlo simulations yield that in the combined inhomogeneous system, the asymmetric region displays coexistence, with more prominent stable spirals, for a larger range of asymmetry factors $k$, as well as a reduced chance of two-species extinction occurring.
This remarkable stabilization phenomenon is realized through the following mechanism: 
The coupling to the symmetric region provides a steady supply of all three population species through periodic invasion fronts emanating from the symmetric-asymmetric interface into the asymmetric region.

Thus, seeding conditions for spiral defects are always present. 
This causes the (quasi-) stationary state of the asymmetric model to feature a healthy set of all three populations, but governed by spatio-temporal features that are similar to the those observed in the fewer surviving realizations of the purely asymmetric model with the same parameters. 
Although coexistence in the coupled case is robust, the eventual evolution of the spirals in the asymmetric subsystem strictly depends on the local parameters in the asymmetric system. 
Furthermore, we have provided a semi-quantitative criterion for these emerging spirals to persist: 
The oscillation frequency of the spirals, determined by the periodic influx from the stable region, needs to be greater than the inverse diffusive transport time scale through the spiral arms, set by the diffusivity multiplied by the square of the inverse of the average correlation length of all three species. 

Two key aspects of the invasions fronts emerging in coupled cyclic population models are worth further discussion here. 
First, we address the variation in the spatio-temporal dynamics with changes in the aspect ratio of the coupled system. 
We have studied the inhomogeneous, coupled symmetric-asymmetric May--Leonard model   for varying system sizes up to $L_x = 2048$, $L_y = 32$. 
As the transverse size $L_y$ is reduced, finite-size effects in this dimension will inevitably constrain pattern formation. 
In our simulation systems, the spiral size is less, but still typically of the order of $L_y$.
The spiral size in the symmetric patch defines a large coherence length along the subsystems' interface, whence the invasion fronts are manifested as essentially planar waves.
Second, in this inhomogeneous coupled symmetric-asymmetric May--Leonard system, we do not discern evidence for a finite penetration depth that would limit the spatial influence of these stabilizing invasion waves.
This is in contrast to other variants of coupled cyclic population models; in particular, for the quite distinct case of a conserved Lotka--Volterra (or ``rock-paper-scissors, RPS'') patch coupled to a symmetric May--Leonard subsystem, one indeed observes a finite penetration depth for the cyclic wavefronts entering the May--Leonard region from the RPS patch~\cite{SSerrao2020}.
\\

\noindent{\bf Acknowledgments:} \\
We are grateful to Michel Pleimling and Priyanka for fruitful discussions and helpful comments.
M. Lazarus Arnau contributed to a preliminary study concerning a distinct system of coupled ecologies, which in part motivated the present work.
This research was sponsored by the Army Research Office and was accomplished under Grant No. W911NF17-1-0156. 
The views and conclusions contained in this document are those of the authors and should not be interpreted as representing the official policies, either expressed or implied, of the Army Research Office or the U.S. Government. The U.S. Government is authorized to reproduce and distribute re\-prints for Government purposes notwithstanding any copyright notation herein. 
\\

\noindent{\bf Author Contribution Statement:} \\
Both authors conceived and planned this study.
S.R.S. wrote the Monte Carlo code, carried out the numerical simulations, performed the quantitative data analysis, and generated the figures.
U.C.T. aided in the interpretation of the results.
Both authors jointly prepared the manuscript.

\bibliographystyle{epj}
\bibliography{references}
\end{document}